\documentclass[a4paper, 11pt]{article}
\usepackage[utf8]{inputenc}
\usepackage{jcappub}
\usepackage{amsfonts}
\usepackage{graphicx}
\usepackage{amssymb}
\usepackage{amsmath}
\usepackage{breqn}
\usepackage{array}
\usepackage{tikz}
\usetikzlibrary{shapes.geometric, arrows}
\tikzstyle{st} = [rectangle, rounded corners, text width = 3cm, text centered, draw = black ]
\tikzstyle{arrow} = [->,>=stealth]

\title{Complete Hamiltonian analysis of cosmological perturbations at
  all orders II: Non-canonical scalar field}

\author{Debottam Nandi} \author{and S. Shankaranarayanan}
\affiliation{School of Physics, Indian Institute of Science Education
  and Research Thiruvananthapuram (IISER-TVM), India}

\emailAdd{debottam@iisertvm.ac.in}
\emailAdd{shanki@iisertvm.ac.in}

\abstract{In this work, we present a consistent Hamiltonian analysis
  of cosmological perturbations for generalized non-canonical scalar
  fields. In order to do so, we introduce a new phase-space variable
  that is uniquely defined for different non-canonical scalar
  fields. We also show that this is the simplest and efficient way of
  expressing the Hamiltonian. We extend the Hamiltonian approach of
  \citep{Nandi:2015ogk} to non-canonical scalar field and obtain
  an unique expression of speed of sound in terms of
    phase-space variable. In order to invert generalized phase-space
  Hamilton's equations to Euler-Lagrange equations of motion, we
  prescribe a general inversion formulae and show that our approach
  for non-canonical scalar field is consistent. We also obtain the
  third and fourth order interaction Hamiltonian for generalized
  non-canonical scalar fields and briefly discuss the extension of our
  method to generalized Galilean scalar fields.}
\begin{document}
\maketitle

\section{Introduction}
The inflationary paradigm is highly successful and an attractive way
of resolving some of the puzzles of standard cosmology. During
inflation, the early universe undergoes an accelerated expansion,
stretching quantum fluctuations to super-horizon scales which we
observe today as CMB anisotropy \cite{Komatsu:2008hk}. Since
Einstein's equations are highly non-linear, comparison of the
predictions of inflation with the observations require one to expand
the equations order-by-order. At linear order, predictions of
inflation is consistent with CMB. However, the linear order
observables, like scalar spectral index and tensor-to-scalar ratio,
can not rule out models of inflation; physically measurable
observables corresponding to higher-order quantities like bispectrum
or trispectrum will help to rule out some models of inflation
\cite{Lidsey:1995np,Lyth:1998xn,Lyth:2007qh,Mazumdar:2010sa}.

In the standard inflationary models, inflation is driven by scalar
field(s). The canonical scalar fields are the simplest and to get
sufficient amount of inflation require flat-potential. During
canonical `slow-roll' evolution, potential energy dominates over its
kinetic energy and drives a quasi-exponential expansion of the
universe which is often difficult to obtain within
  particle physics models \cite{Lyth:1998xn,
    Lidsey:1995np}. Non-canonical scalar fields are generalizations
of canonical scalar fields and reduces the dependence on the
potential. In case of non-canonical scalar field, even in the absence
of potential energy term, a general class of non-quadratic kinetic
terms can drive inflationary evolution. This model satisfy two crucial
requirements of inflationary scenarios: the scalar perturbations are
well-behaved during inflation and there exists a natural mechanism for
exiting inflation in a graceful manner. Non-canonical scalar field
also contains extra parameters than canonical scalar field such as
speed of sound. However, unlike canonical scalar field modelds, the
speed of propagation of the scalar perturbations in these inflationary
models can be time-dependent\cite{Armendariz-Picon1999, Garriga1999,
  Armendariz-Picon2001}. Recently, in order to seek more generalized
field, scalar fields with higher time derivatives models like
Hordenski scalar fields, Kinetic Gravity Braiding models \cite{Horndeski1974,Kobayashi2010,Kobayashi2011,Nicolis2008,Deffayet2009,Deffayet:2010qz}
are considered. Besides these, there are plenty of other models that
lead to accelerated universe.

Since inflation takes place at high-energies, quantum field theory is
the best description of the matter at these energies.  Hence,
evaluation of any physical quantity, like the $n-$point correlation
functions, require us to either promote effective field variables
(using Heisenberg picture) to operators or integrate over all possible
field configurations on all of space–time (path - integral
picture). Since it is unclear what effective field configurations to
integrate over, Heisenberg picture is the preferred approach. In other
words, we obtain the effective Hamiltonian operator and evaluate the
correlation functions of the relevant operator.

In the context of cosmological perturbation theory, there are
currently two approaches in the literature to evaluate the effective
Hamiltonian --- Lagrangian fomalism and Hamiltonian formalism. In case
of Lagrangian formalism,  the Lagrangian is expanded upto
  a particular order, i.e., if we are interested in obtaining third
  order interaction Hamiltonian, effective Lagrangian needs to be
  expanded up to third order and constraints are systematically
  removed from the system to obtain the effective perturbed
  Lagrangian. Then, the momentum $\pi$ corresponding to $\varphi$ is
  obtained as a polynomial of $\dot{\varphi}$ and using order-by-order
  approximations, $\dot{\varphi}$ is expressed as a polynomial of
  $\pi$. Next, using Legendre transformation, Hamiltonian is expressed
  in terms of $\pi$ and $\varphi$. In order to express the Hamiltonian
  in terms of $\dot{\varphi}$ and $\varphi$, only the leading order
  relation between $\pi$ and $\dot{\varphi}$ is used
  \cite{Huang:2006eha, Seery:2005wm, Chen:2006nt, Bartolo:2001cw,
    Bartolo:2003bz, Bartolo:2004if, Maldacena2003, Seery:2008ax}. There are some difficulties with
  the previous method:
\begin{enumerate}
\item In case of cosmological perturbations, $\pi$ and $\dot{\varphi}$
  are perturbed quantities (curvature perturbation), expressing one in
  terms of polynomial of other is an approximation.
\item At the end, to express the Hamiltonian in terms of
  configuration-space variable, we use only the leading order relation
  between $\pi$ and $\dot{\varphi}$, not the polynomial
  relation. Hence, several approximations are employed to convert
  effective Lagrangian to effective Hamiltonian.
\item The above method is also very restrictive and it is 
  difficult to extend the method for a generalized constrained
  system.
\item Also, it is difficult to use this method for higher
  order perturbations.
\end{enumerate}

In the context of cosmological pertubrations, the above approach leads to consistent results, however, a consistent Hamiltonian formulation is always preferred than the previous approach to make calculations simpler with more technical details. Langolois \cite{Langlois1994} first introduced a consistent
Hamiltonian formulation of canonical scalar field. However, Langlois'
approach is also difficult to extend to higher order of perturbations
or to any different types of field due to the fact that it requires
construction of gauge-invariant conjugate momentum. In our recent work
\cite{Nandi:2015ogk}, henceforth referred as I, we have introduced a
different Hamiltonian approach that can address and deal with all the
issues in the previous methods and provides an effective and robust
way to obtain interaction Hamiltonian for any model for any order of
perturbations. Also, in case of calculating mixed-mode (e.g.,
  scalar-tensor) interaction Hamiltonian \cite{Maldacena2003, Seery:2008ax}, our approach is simpler than the previous one. Table below provides a bird's eye view of the both the
formulations and advantages of the Hamiltonian formulation that is
proposed in this work \cite{Nandi:2015ogk}

\begin{center}
\begin{tabular}{|m{2cm}|m{6cm}|m{6cm}|}
  \hline
  & Lagrangian formulation & Hamiltonian formulation \\
  \hline
  Gauge conditions and {guage-invariant equations} & At any order, choose a 
gauge {which} does not lead to gauge-artifacts & 
  Choose a gauge {with no gauge-artifacts}, however, momentum corresponding 
to unperturbed quantity is non-zero leading to consistent equations of motion.\\
  \hline
  Dynamical variables & {Counting true dynamical degrees of freedom is 
difficult.} & Using Dirac's procedure, constraints can 
  easily be obtained and is easy to determine the degrees of freedom.\\
  \hline
  Quantization at all orders &  Difficult to quantize constrained systems. & 
Since constraints are obtained systematically and reduced phase space contains 
only true degrees of freedom, it is straightforward to quantize the theory 
using Hamiltonian formulation.\\
  \hline
  Calculating the observables & Requires to invert the expressions at each order and hence non-trivial 
to compute higher-order correlation function. & Once the relation between 
$\varphi$ and Curvature perturbation\footnotemark[1]  is known, 
calculating the correlation functions from the Hamiltonian is simple and straightforward to obtain.\\
  \hline
\end{tabular}
\end{center}

\footnotetext[1]{It is important to note that, in the case of first
  order, relation between $\varphi$ and three-curvature is straight
  forward. However, it is more subtle in the case of higher-order
  perturbations\cite{Malik2009}.}
  
 In I, we applied our new method to canonical and a specific higher derivative (Galilean) scalar field model, and showed
explicitly that the method can efficiently obtain Hamiltonian at all
orders. In the case of certain non-canonical scalar field models, if
$\dot{\varphi}$ can be expressed uniquely in terms of the canonical
conjugate momentum, it is then possible to obtain Hamiltonian and the
results of I can be extended. However, for a general non-canonical
scalar field, it is not possible to do the procedure as we do not have
a way to rewrite $\dot{\varphi} $ in terms of the canonical conjugate
momentum and, hence, it is not possible to obtain the Hamiltonian for
general non-canonical scalar field. In this work, we explicitly obtain
Hamiltonian for a general non-canonical scalar fields and obtain
interaction Hamiltonian upto fourth order.

This work is divided into two parts. In the first part, we provide the
procedure to obtain Hamiltonian for the non-canonical scalar field by
introducing a phase-space variable. Then, by choosing different
models, we explicitly show that the Hamiltonian leads to consistent
equations of motion as well as perturbed interaction Hamiltonian by
implementing our approach. We also find a new definition of speed of
sound in terms of phase-space variables. In the second part, in order
to retrieve generalized equations of motion in configuration-space
from phase-space, we provide a systematic way to invert generalized
non-canonical phase-space variables to configuration-space variables
and vice versa and show that, all equations are consistent. Finally,
we extend the method to generalized higher derivative
  scalar fields. A flow-chart below illustrates the method for
non-canonical scalar fields:

{\tiny
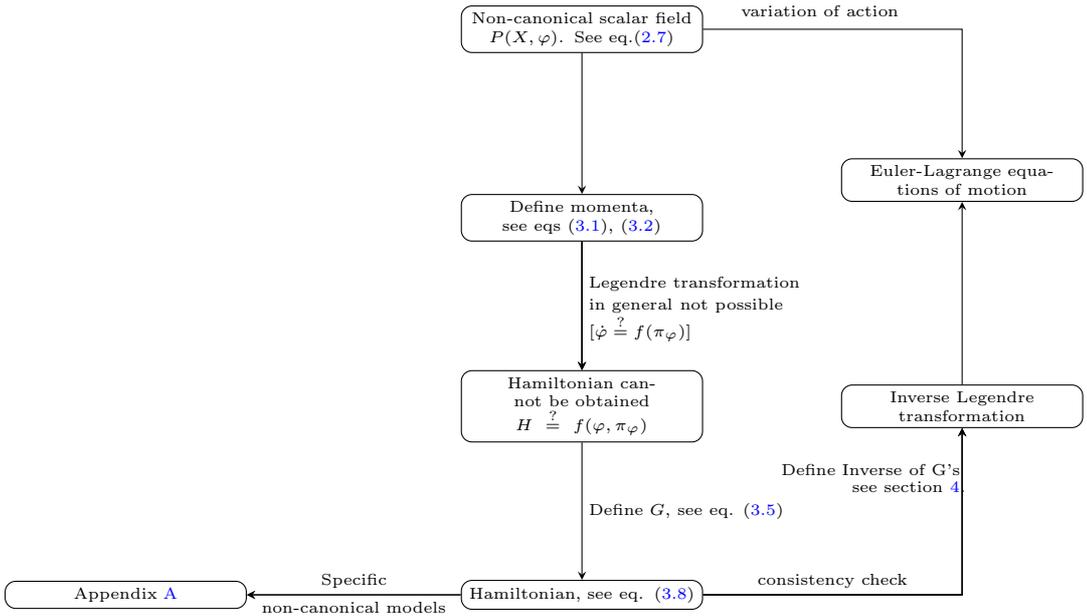
\begin{center}
\begin{tikzpicture}[node distance=2.5cm]

  \node (start)[st] {Non-canonical scalar field \\ $P(X,
    \varphi)$. See eq.(\ref{ADM-Non-canonical})}; \node (defmom) [st,
  below of=start] {Define momenta, see eqs (\ref{piijgeneral}),
    (\ref{piphigeneral})}; \node (nonham) [st, below
  of=defmom]{Hamiltonian cannot be obtained\\$H \stackrel{?}{=}
    f(\varphi, \pi_\varphi) $}; \node (ham) [st, below of=nonham
  ]{Hamiltonian, see eq. (\ref{Hamfull})}; \node (Appndix) [st, below
  of=nonham, xshift= -6cm] {Appendix \ref{Inversion}}; \node (Inv)
  [st, below of=defmom, xshift=5cm]{Inverse Legendre transformation};
  \node (EL) [st, below of=ham, yshift=8cm, xshift=5cm]{Euler-Lagrange
    equations of motion};

\draw [arrow] (start) -- (defmom);
\draw [arrow] (defmom) --node[anchor=west, yshift=.3cm]{Legendre transformation} (nonham);
\draw [arrow] (defmom) --node[anchor=west]{in general not possible} (nonham);
\draw [arrow] (defmom) --node[anchor=west, yshift=-.3cm]{$[\dot{\varphi} \stackrel{?}{=} f(\pi_\varphi)$]} (nonham);
\draw [arrow] (nonham) -- node[anchor=west]{Define $G$, see eq. (\ref{G})}(ham);
\draw [arrow] (ham) -- node[anchor=south]{Specific}(Appndix);
\draw [arrow] (ham) -- node[anchor=north]{non-canonical models}(Appndix);
\draw [arrow] (ham) -| node[anchor=south, xshift=-1.7cm]{consistency check} (Inv);
\draw [arrow] (ham) -| node[anchor=north, xshift=-.7cm, yshift=1.6cm]{see section \ref{InversionNC}.} (Inv);
\draw [arrow] (ham) -| node[anchor=south, xshift=-1.2cm, yshift=1.5cm]{Define Inverse of G's} (Inv);
\draw [arrow] (start) -| node[anchor=west, xshift=-3cm, yshift=.25cm]{variation of action} (EL);
\draw [arrow] (Inv) -- (EL);

\end{tikzpicture}
\end{center}
}

In the next section, we briefly discuss about non-canonical scalar
fields. We also discuss the gauge choices and corresponding
gauge-invariant variables. In section \ref{Hamiltonian-nonC},
Hamiltonian formulation of generalized non-canonical scalar field is
introduced by defining a new phase-space function which provides
consistent equations of motion. In section \ref{first}, we extend the
results of I to non-canonical scalar field in flat-slicing gauge to
obtain perturbed equations of motion. In section \ref{InversionNC}, we
provide a partial inversion method between phase-space variables and
configuration-space variables and in section \ref{InteractionHam}, we
provide the third and fourth order Interaction Hamiltonian for
non-canonical scalar field. In section \ref{ExGal}, we briefly discuss
the application of our method to generalized Galilean scalar field
model. Finally, in section \ref{Conclu}, we end with discussions and
conclusions of the results. In Appendix \ref{Inversion}, functional
form of the new variable is obtained for different scalar field
models. In Appendix \ref{Langlois-nonC}, we implement Langlois'
approach to non-canonical scalar field model.

In this work, we consider $(-, + , +, +)$ metric signature. We also
denote $\prime$ as derivative with respect to conformal time.

\section{Model and gauge choices}\label{BasicModel}
Action for non-canonical scalar field minimally coupled
to gravity is
\begin{equation}
\label{GravityAction}
\mathcal{S} = \int d^4 x \sqrt{-g} \left[\frac{1}{2 \kappa} R + \mathcal{L}_m(\varphi, \partial\varphi) \right],
\end{equation}
where $R$ is the Ricci scalar and the matter Lagrangian, $\mathcal{L}_m$ is of the form.
\begin{equation}\label{boxx}
  \mathcal{L}_m = P(X, \varphi)  ,~~~ ~X \equiv 
  \frac{1}{2} g^{\mu \nu} \partial_{\mu}{\varphi} \partial_{\mu}{\varphi}.
\end{equation}

{$P(X, \varphi)$ corresponds to non-standard kinetic term
  and hence the name non-canonical scalar
  field\cite{Armendariz-Picon1999,Garriga1999,Armendariz-Picon2001}. Further,
  fixing $ P = - X - V(\varphi)$, where $V(\varphi)$ is the potential,
  one can retrieve the well-known canonical scalar field model.}
Varying the action (\ref{GravityAction}) with respect to metric gives
Einstein's equation
\begin{equation}
\label{EinsteinEquation}
R_{\mu \nu} - \frac{1}{2} g_{\mu \nu}~ R = \kappa~ T_{\mu \nu}, 
\end{equation}
where the stress tensor $(T_{\mu \nu})$ for non-canonical scalar field is
\begin{equation}
\label{EMTensor}
T_{\mu \nu} = - P_X \partial_{\mu}{\varphi} \partial_{\nu}{\varphi} + g_{\mu \nu}  P.
\end{equation}

Varying the action (\ref{GravityAction}) with respect to the scalar
field `$\varphi$' leads to the following equation of motion
\begin{equation}
\label{EG}
   P_X \Box \varphi - P_{XX} \partial_{\mu}{\varphi} \partial^{\mu}{\varphi} -
  2X
   P_{X \varphi}  + P_\varphi = 0
\end{equation}
which can also be obtained from the conservation of Energy-Momentum
tensor, $\nabla_{\mu}{T^{\mu \nu}} = 0$.

The four-dimensional line element in the ADM form is given by,
\begin{eqnarray}
ds^2 &=& g_{\mu \nu} dx^{\mu} dx^{\nu} \nonumber \\
\label{line}
&=& -(N^2 - N_{i} N^{i} ) d\eta^2 + 2 N_{i} dx^{i} d\eta + \gamma_{i j} dx^{i} dx^{j},
\end{eqnarray}
where $N(x^{\mu})$ and $N_i(x^{\mu})$ are Lapse function and Shift
vector respectively, $\gamma_{i j}$ is the 3-D space metric.  Action
(\ref{GravityAction}) for the line element (\ref{line}) takes the form
\begin{equation}
\label{ADM-Non-canonical}
\mathcal{S}_{NC} = \int d^4x\, N\, \sqrt{\gamma}\, \left[ \frac{1}{2 \kappa}
\left(^{(3)}R + K_{i j} K^{i j} - K^2\right) + P(X, \varphi) \right]
\end{equation}
where $K_{i j}$ is extrinsic curvature tensor and is defined by
\begin{eqnarray}
  && K_{i j} \equiv \frac{1}{2N} \left( \partial_{0}{\gamma_{i j}} - N_{i|j} - N_{j | i} \right), \nonumber \\
  && K \equiv \gamma^{i j} K_{i j}. \nonumber
\end{eqnarray}

Perturbatively expanding the metric only in terms of scalar
perturbations and the scalar field about the flat FRW spacetime in
conformal coordinate, we get,

\begin{eqnarray}
  && g_{0 0} = - a(\eta)^2(1 + 2 \epsilon \phi_1 + \epsilon^2 \phi_2 + ...) \\
  && g_{0 i} \equiv N_{i} = a(\eta)^2 (\epsilon \partial_{i}{B_1} + \frac{1}{2} \epsilon^2 \partial_{i}{B_2} + ...) \\
  && g_{i j} =a(\eta)^2 \left((1 - 2 \epsilon \psi_1 - \epsilon^2 \psi_2 -...)\delta^{i j} +  2 \epsilon E_{1 i j} + \epsilon^2 E_{2 i j} + ...\right)\\
  &&\varphi = \varphi_0(\eta) + \epsilon \varphi_1 + \frac{1}{2} \epsilon^2 \varphi_2 + ...
\end{eqnarray}
where $\epsilon$ denotes the order of the perturbation.  To determine
the dynamics at every order, we need five scalar functions ($\phi, B,
\psi, E$ and $\varphi$) at each order. Since there are two arbitrary
gauge-freedoms for scalar perturbations, one can fix two of the five
scalar functions. In this work, we derive all equations by choosing a
specific gauge --- flat-slicing gauge, i.e., $\psi = 0, E = 0$ --- at
all orders:
\begin{eqnarray}
\label{00pmetric}
  && g_{0 0} =- a(\eta)^2(1 + 2 \epsilon \phi_1 + \epsilon^2 \phi_2 + ...) \\
  \label{0ipmetric}
  && g_{0 i} \equiv N_{i} = a(\eta)^2 (\epsilon \partial_{i}{B_1} + \frac{1}{2} \epsilon^2 \partial_{i}{B_2} + ...) \\
  \label{3metric}
  && g_{i j} =a(\eta)^2 \delta_{i j}\\
  \label{field}
  &&\varphi = \varphi_0(\eta) + \epsilon \varphi_1 + \frac{1}{2} \epsilon^2 \varphi_2 + ...
\end{eqnarray}

It can be shown that, perturbed equations in flat-slicing gauge
coincide with gauge-invariant equations of motion (in generic gauge,
$\varphi_1$ coincides with $\varphi_1 + \frac{\varphi_0{}^\prime}{H}
{}\psi_1 \equiv \frac{\varphi_0{}^\prime}{H} \mathcal{R}$ which is a
gauge-invariant quantity, $\mathcal{R}$ is called curvature
perturbation \cite{Malik2009}). Similarly, one can choose another
suitable gauge with no coordinate artifacts to obtain gauge-invariant
equations of motion. Such gauges are Newtonian-conformal gauge $(B =
0,\, E = 0)$, constant density gauge $(E = 0, \, \delta\varphi = 0)$,
etc.
  
Before we proceed with the Hamiltonian formulation, it is important to
clarify issues related to the quantization in the cosmological
perturbation theory: While the field variable $\varphi_i$ and metric
variables $\phi_i, B_i, \psi_i$ (where $i$ takes values $1, 2,
\cdots$) are expanded perturbatively, it is important to note that the
operators corresponding to these variables (i. e. $\varphi_1,
\varphi_2, \cdots$) can not be treated as independent operators as
higher orders in perturbation theory do not lead to independent
degrees of freedom. As otherwise, the unperturbed theory should have
infinitely many local degrees of freedom.  In a canonical
quantization, there is one operator and one for its momentum, on which
the quantum Hamiltonian depend\footnote{We thank Martin Bojowald for
  discussion regarding this point.}.

\section{Hamiltonian formulation}\label{Hamiltonian-nonC}

In our last work \cite{Nandi:2015ogk}, we provided an efficient way of
obtaining consistent perturbed Hamiltonian for any gravity
models. However, it works only if the form of the action is specified
in such a way that the Legendre transformation $\dot{\varphi}
\rightarrow \pi_\varphi$ is invertible in both ways. For non-canonical
scalar fields, momenta corresponding to the action
(\ref{ADM-Non-canonical}) are
\begin{eqnarray}
\label{piijgeneral}
\pi^{i j} \equiv \frac{\delta \mathcal{S}_{NC}}{\delta {\gamma^\prime_{i j}}}  &=& \frac{\sqrt{\gamma}}{2 \kappa}\,  ({\gamma}^{i j} {\gamma}^{k l} - {\gamma}^{i k} {\gamma}^{j l}) {K}_{k l} \\
 \label{piphigeneral}
 \pi_\varphi \equiv \frac{\delta \mathcal{S}_{NC}}{\delta {\varphi^\prime}} &=& - \sqrt{\gamma}P_X \sqrt{- 2X +  Y}, \quad \mbox{where}\quad Y \equiv \gamma^{i j} \partial_i \varphi \partial_j \varphi.
\end{eqnarray}
% $\pi_\varphi \equiv \pi_\varphi ( X, \gamma, Y, \varphi)$ and $X \equiv X (\pi_\varphi, \gamma, Y, \varphi)$. 

As one can see, equation (\ref{piijgeneral}) is invertible and the
inversion relation is given by
\begin{equation}\label{InvPiij}
 {\gamma}^\prime_{m n} = \gamma_{n k} N^k_{|m} + \gamma_{m k} N^k_{|n} - 2 N K_{m n}, ~~~ K_{i j} = \frac{\kappa}{\sqrt{\gamma}} \left(\gamma_{i j} \gamma_{k l} - 2 \gamma_{i k} \gamma_{j l}\right)
\pi^{k l} 
\end{equation}
but equation (\ref{piphigeneral}) is non-invertible for arbitrary
function of $P(X, \varphi)$. However, if $P(X, \varphi)$ is specified,
it may be possible to invert the equation and $X$ can be written in
terms of $\pi_\varphi$. Inversion relations for commonly used
non-canonical models are given in Appendix \ref{Inversion}. Using
equation (\ref{InvPiij}), we can write the Hamiltonian density as
\begin{eqnarray}
  \mathcal{H}_{NC} &=& \pi^{i j} \gamma_{i j}^\prime + \pi_\varphi \varphi^\prime - \mathcal{L}_{NC} \nonumber \\
  &=& 2 {\gamma}_{i j} {\partial}_{k}{{N}^{j}}\,  {\pi}^{i k} + {N}^{i} {\partial}_{i}{{\gamma}_{j k}}\,  {\pi}^{j k}  - \frac{N \kappa}{\sqrt{\gamma}} \left(\gamma_{i j} {\gamma}_{k l} - 2\gamma_{i k} \gamma_{j l}\right) {\pi}^{i j} {\pi}^{k l}  - \frac{N \sqrt{\gamma}}{2 \kappa} \,{}^{(3)}R  -\nonumber \\
 \label{fullhamiltonian}
 &&  N \sqrt{\gamma}~\tilde{G}(X, \gamma, Y, \varphi) + {N}^{i} \pi_\varphi {\partial}_{i}{\varphi}, \quad\mbox{where}\quad\tilde{G}\equiv \left( P - P_X \left(2X - Y\right)\right).
\end{eqnarray}

Note that, the above expression is still not the Hamiltonian since
$\tilde{G}$ is not a phase-space variable and it is not invertible for
arbitrary form of $P(X, \varphi)$ since equation (\ref{piphigeneral})
is not invertible, in general. Hence, a natural question that arises
is: \emph{How to invert configuration-space variables to phase-space
  variables so that we can obtain generalized consistent Hamiltonian
  for non-canonical scalar field?}
 
In this section, we show that, by defining a new phase-space function,
the above problem can be resolved. The new phase-space quantity is
defined as
\begin{equation}
\label{G}
G(\pi_\varphi, \gamma, Y, \varphi) = \tilde{G}(X, \gamma, Y, \varphi) \equiv P - P_X \left(2X - Y\right).
\end{equation}

Since momenta corresponding to $N$ and $N^i$ vanish, i.e., $\pi_N =
\pi_i = 0$, using the above defined variable, Hamiltonian constraint
can be written as
\begin{equation}
\label{HamitonianConstraint}
\mathcal{H}_N \equiv \{\pi_N, \mathcal{H}_{NC}\} = \frac{\delta \mathcal{H}_{NC}}{\delta N} = - \frac{ \kappa}{\sqrt{\gamma}} \left(\gamma_{i j} {\gamma}_{k l} - 2\gamma_{i k} \gamma_{j l}\right) {\pi}^{i j} {\pi}^{k l}  - \frac{ \sqrt{\gamma}}{2 \kappa} \,{}^{(3)}R  -  \sqrt{\gamma}\,\, G(\pi_\varphi, \gamma, Y, \varphi) = 0,
\end{equation}
and Momentum constraint is given by
\begin{equation}
\label{MomentumConstraint}
M_i \equiv \{\pi_i, \mathcal{H}_{NC}\} = \frac{\delta \mathcal{H}_{NC}}{\delta N^i}= -2 \partial\left(\gamma_{i m}\pi^{m n}\right) + \pi^{k l} \partial_{i}\gamma_{k l} + \pi_\varphi\partial_i \varphi = 0.
\end{equation}

Due to diffeomorphic invariance, total Hamiltonian density can be written as
\begin{equation}\label{Hamfull}
\mathcal{H}_{NC} = N \mathcal{H}_N + N^i \mathcal{H}_i = 0. 
\end{equation}

Instead of defining $G$, one can define any other phase-space
variable(s) and express the Hamiltonian in a different form and the
possibilities are infinite. However, as one can see, since
$\tilde{G}(X, \gamma, Y, \varphi)$ automatically appears directly in
the Hamiltonian, this is the simplest and effective way to express the
Hamiltonian for non-canonical scalar field. $G$ not only resolves the
issue of expressing Hamiltonian for non-canonical scalar field and is
also uniquely defined for different non-canonical scalar
fields. Hence, $G$ carries the signature of the non-canonical scalar
fields in phase-space. Explicit forms of $G(\pi_\varphi, \gamma, Y,
\varphi)$ for different types of scalar fields are given in Appendix
\ref{Inversion}.

\subsection{Zeroth order}\label{zero}
At zeroth order, since $\gamma_{i j} = a^2 \delta_{i j}$ and all
quantities are independent of spatial coordinates, we then get
\begin{equation}
\pi_0^{i j} = \frac{1}{6 a} \pi_a \delta^{i j} 
\end{equation}
and Hamiltonian density at zeroth order becomes
\begin{equation}
\label{zerothH}
\mathcal{H}_{0} = - \frac{N_0 \kappa}{12 a} \pi_a^2 - G \,N_0\, a^3.
\end{equation}

Variation of the Hamiltonian (\ref{zerothH}) with respect to the
momenta leads to two equations and are given by
\begin{eqnarray}
\label{Zerotha'}
a^\prime &=& - \frac{N_0 \kappa}{6 a} \pi_a \\
\label{Zerothphi'}
\varphi_0^\prime &=& - N_0 \,a^3\, G_{ \pi_\varphi}.
\end{eqnarray}

Hamiltonian constraint leads to the equation of motion of $N$ and at
zeroth order, it is given by
\begin{equation}\label{ZerothHam}
\mathcal{H}_{N0}  \equiv - \frac{\kappa}{12 a} \pi_a^2 - G\, a^3 = 0.
\end{equation}

Equations of motion are obtained by varying the Hamiltonian
(\ref{zerothH}) with respect to field variables. Hence, equation of
motion of $a$ is obtained by the relation
\begin{equation}\label{ZerothEoMa}
  \pi_a^\prime = - \frac{\delta \mathcal{H}_{0}}{\delta a} = - \frac{N_0 \kappa}{12 a^2} \pi_a^2 + 3\, G\, N_0\, a^2 + G_{a}\, N_0\, a^3.
\end{equation}

Similarly, equation of motion of $\varphi_0$ can be obtained from
\begin{equation}\label{ZerothEoMphi}
\pi_{\varphi0}^\prime = N_0\, a^3\,G_{\varphi}.
\end{equation}

\subsection{First order}\label{first}

As we have mentioned in the introduction, there are two ways to obtain
Hamiltonian --- Langlois' approach \citep{Langlois1994}, and the
approach used in I \cite{Nandi:2015ogk}. In this work, we use both the
approaches and explicitly show that it is possible to obtain a
consistent Hamiltonian for non-canonical scalar fields. In Appendix
\ref{Langlois-nonC}, we extend Langlois' approach to non-canonical
scalar field and in the rest of the section, we extend I to obtain a
consistent Hamiltonian for non-canonical scalar field.

The field variables and their corresponding momenta can be separated
into unperturbed and perturbed parts as
\begin{eqnarray}
\label{perturbfm1}
&& N = N_0 + \epsilon N_1, ~~~N^{i} = \epsilon N_1^i,~~~\varphi = \varphi_0 + \epsilon \varphi_1 \\
\label{perturbfm2}
&& \pi^{i j} = {}\pi_0{}^{i j} + \epsilon \pi_1^{i j}, ~~~\pi_\varphi = {}\pi_{\varphi0} + \epsilon \pi_{\varphi1}
\end{eqnarray}
and by using Taylor expansion of phase-space variable $G(\pi_\varphi,
\gamma, Y, \varphi)$, the second order perturbed Hamiltonian density
is given by
\begin{eqnarray}
\label{firstH}
&& \mathcal{H}_{2} =  
{\delta}_{i j} {\partial}_{k}{{N_1}^{j}}\,  ({\pi_1}^{i k} + \pi_1^{k i}) a{}^{2}  - N_0  \kappa a\,({\delta}_{i j} {\delta}_{k l} - 2 \delta_{i k} \delta_{j l}){\pi_1}^{i j} {\pi_1}^{k l} - 2\, N_1 \kappa a \,({\delta}_{i j} {\delta}_{k l} - 2 \delta_{i k} \delta_{j l}){\pi_0}^{i j} {\pi_1}^{k l} \nonumber \\
&&~~~~~ - G_\varphi N_1 a{}^{3} \varphi_1 - G_{\pi_\varphi} N_1 \,\pi_{\varphi 1} \,a{}^{3} - \frac{1}{2}\, G_{\varphi\varphi} N_0\, \varphi_1{}^{2} a{}^{3} - \frac{1}{2}\, G_{\pi_\varphi \pi_\varphi} N_0\, \pi_{\varphi 1} {}^{2} a{}^{3} - G_{\varphi \pi_\varphi } N_0 \pi_{\varphi 1} a{}^{3} \varphi_1 \nonumber \\
&&~~~~~- G_Y N_0 {\delta}^{i j} {\partial}_{i}{\varphi_1}\,  {\partial}_{j}{\varphi_1}\,  a + {N_1}^{i} \pi_{0 \varphi} {\partial}_{i}{\varphi_1}.\,
\end{eqnarray}

Note that, as we have pointed out in I \citep{Nandi:2015ogk},
perturbed momentum corresponding to an unperturbed variable may arise
due to the presence of other perturbed phase-space variables, thus
$\pi^{ij}_1$ is non-zero and can be obtained by varying the
Hamiltonian (\ref{firstH}) with respect to $\pi^{ij}_1$:
\begin{equation}
  \frac{\delta \mathcal{H}_2}{\delta \pi_1^{i j}} = 0 \Rightarrow \pi^{i j}_1 = \frac{a}{2 N_0 \kappa} \delta^{i j} \partial_k N_1^k - \frac{a}{4 N_0 \kappa} \delta^{k i} \partial_k N_1^j - \frac{a}{4 N_0 \kappa} \delta^{k j} \partial_k N_1^i - \frac{N_1}{N_0} \pi_0^{i j}.
\end{equation}

Varying the perturbed Hamiltonian (\ref{firstH}) with respect to $\pi_{\varphi1}$ leads to the following equation
\begin{eqnarray}
  \varphi_1^\prime &=& - G_{\pi_\varphi} N_1 a^3 - G_{\pi_\varphi \pi_\varphi}N_0\, \pi_{\varphi1} a^3 - G_{\varphi \pi_\varphi} N_0\, \varphi_1 a^3 \\
  \Rightarrow \pi_{\varphi1} &=& - \frac{1}{N_0\, a^3\,G_{\pi_\varphi \pi_\varphi} }\left(\varphi_1^\prime + G_{\pi_\varphi} N_1 a^3 + G_{\varphi \pi_\varphi} N_0\, \varphi_1 a^3 \right).
\end{eqnarray}

Hamiltonian constraint is obtained by varying the Hamiltonian with
respect to Lapse function. Hence, varying (\ref{firstH}) with respect
to $N_1$ leads to first order Hamiltonian constraint and takes the
form
\begin{equation}
\label{FirstHamiltonian}
-2 \delta_{i j} \delta_{k l} \,\pi_0^{i j}\, \pi_1^{k l} + 4 \delta_{i j} \delta_{k l} \,\pi_0^{i k}\, \pi_1^{j l} - G_\varphi a^2 \varphi_1 - G_{\pi_\varphi} \pi_{\varphi1}\, a^2 = 0.\\
\end{equation}

Similarly, by varying Hamiltonian with respect to $N_1^i$ we get the following Momentum constraint,
\begin{equation}
\label{FirstMomentum}
\pi_{\varphi0} \, \partial_i \varphi_1 - 2 a^2\, \delta_{i j}\,\partial_k  \pi_1^{k j}=0.
\end{equation}

Finally, equation of motion of $\varphi_1$ is obtained by varying the Hamiltonian with respect to $\varphi_1$, i.e.,
\begin{equation}
\label{FirstEoMvarphi}
\pi_{\varphi_1}^\prime = a^3 \,G_\varphi N_1  + a^3\,G_{\varphi\varphi}N_0 \varphi_1  + a^3\,G_{\varphi \pi_\varphi} N_0 \pi_{\varphi1} - 2\,a\, G_Y N_0 \nabla^2 \varphi_1 + \pi_0^{i j}\,\partial_iN_1^i.
\end{equation}

Since the perturbed scalar field equation is linear in nature and
follows wave equation, speed of sound is defined as the ratio of
negative of the coefficient of $\nabla^2 \varphi_1$ and
$\varphi_1^{\prime\prime}$ and in phase-space, it takes the form
\begin{equation}
c_s^2 = 2 \,N_0^2\,a^4 \, G_{\pi_\varphi \pi_\varphi}\, G_Y
\end{equation}
which, in conformal coordinate can be expressed as
\begin{equation}\label{sonic-conf}
c_s^2 =  2 \,a^6 \, G_{\pi_\varphi \pi_\varphi}\, G_Y.
\end{equation}

The relation between generalized phase-space derivatives of $G$
($G_\varphi,~ G_Y,~G_{\varphi \pi_\varphi}$ etc.) and
configuration-space derivatives of $P(X, \varphi)$
($P,~P_{\varphi},~P_{\varphi X}$ etc.) is unknown, hence, it is not
possible to invert above Hamilton's equations to Euler-Lagrange
equations and hence, it is not possible to compare both the
formalisms. However, for a particular scalar field, the exact form of
$G$ is known to us (see Appendix \ref{Inversion}), and hence, for
those model it is possible to write down equations of motion in
configuration space and can be verified that Hamiltonian formulation
of non-canonical scalar field is consistent.

\section{Inversion of non-canonical terms}\label{InversionNC}
In the last section, we showed that it is possible to obtain
Hamiltonian for a non-canonical scalar field by defining a new
variable G (see eqs. (\ref{G}) and (\ref{Hamfull})).  In order to
understand the importance of this new function G, we ask the following
question: \emph{Starting from the Hamiltonian (\ref{zerothH}) and
  (\ref{firstH}), can we invert the expressions leading to generalized
  equations in configuration-space?} In this section, we show that,
inversion can be established.

To invert the equations, one needs to invert the coefficients like
$G_\varphi,~ G_Y,~G_{\varphi \pi_\varphi}$ from phase-space to
configuration-space. Since the form of $G$ in configuration space is
known, by carefully looking at the equations, it is apparent that only
the phase-space derivatives of $G$ are needed to invert which, in
general, is not possible.

To begin with, let us take a phase-space function $F \equiv
F(\pi_\varphi, \gamma, Y, \varphi) = \tilde{F}(X, \gamma, Y,
\varphi)$, i.e.,
\begin{eqnarray}
 F &=& F(\pi_\varphi, \gamma, Y, \varphi) \nonumber \\
\Rightarrow dF &=& F_{\pi_\varphi} d\pi_\varphi + F_\gamma d\gamma + F_Y  dY + F_\varphi d\varphi \nonumber
\end{eqnarray}

Note that, tilde is used for configuration-space functions only. The invertibility of Legendre transformation implies that, if $X=X(\pi_\varphi, \gamma, Y, \varphi)$ then $\pi_\varphi = \pi_\varphi (\pi_\varphi, \gamma, Y, \varphi)$, i.e., 

\begin{eqnarray}
&& \pi_\varphi = \pi_\varphi ( X, \gamma, Y, \varphi) \nonumber \\
&& d\pi_\varphi = \frac{\partial \pi_\varphi}{\partial X} dX + \frac{\partial \pi_\varphi}{\partial \gamma}d\gamma + \frac{\partial \pi_\varphi}{\partial Y} dY + \frac{\partial \pi_\varphi}{\partial \varphi} d \varphi \nonumber 
\end{eqnarray}
implying that
\begin{eqnarray}
 d\tilde{F}(X, \gamma, Y, \varphi) &=& F_{\pi_\varphi} \frac{\partial \pi_\varphi}{\partial X} dX + \left( F_\gamma + F_{\pi_\varphi} \frac{\partial \pi_\varphi}{\partial \gamma}\right)  d\gamma + \left( F_Y + F_{\pi_\varphi}\frac{\partial \pi_\varphi}{\partial Y}\right)  dY \nonumber \\
&&+\left( F_\varphi +  F_{\pi_\varphi}\frac{\partial \pi_\varphi}{\partial \varphi}\right) d\varphi.
\end{eqnarray}

Hence, the relations between phase-space variables and
configuration-space variables are

\begin{eqnarray}
 F_{\pi_\varphi} &=& \frac{\tilde{F}_X}{\frac{\partial \pi_\varphi}{\partial X}}, \\
 F_\gamma &=& \tilde{F}_\gamma -  F_{\pi_\varphi} \frac{\partial \pi_\varphi}{\partial \gamma}, \\
 F_Y &=& \tilde{F}_Y -  F_{\pi_\varphi} \frac{\partial \pi_\varphi}{\partial Y},\\
 F_\varphi &=& \tilde{F}_\varphi -  F_{\pi_\varphi} \frac{\partial \pi_\varphi}{\partial \varphi}.
\end{eqnarray}

In our case, for arbitrary non-canonical scalar field, we do not know
the exact form of $G(\pi_\varphi, \gamma, Y, \varphi)$, however, we
know $\tilde{G}(X, \gamma, Y, \varphi) = P - P_X \left(2X -
  Y\right)$. Using equation (\ref{piphigeneral}) and the above
established relations, we get

\begin{eqnarray}
&&G_{\pi_\varphi} =  - \frac{\sqrt{-2X}}{a^3},~~~G_{\pi_\varphi \pi_\varphi} = \frac{1}{a^6 \left(P_X + 2X P_{XX}\right)},\nonumber \\
 &&G_{\pi_\varphi \pi_\varphi \pi_\varphi} = -\frac{\sqrt{-2X} \left(3P_{XX} + 2X P_{XXX}\right)}{a^9 \left(P_X + 2X P_{XX}\right)^3} \nonumber \\
 &&G_\varphi = P_\varphi, ~~~~~G_{\varphi \pi_\varphi} = \frac{\sqrt{-2X}\,P_{X\varphi}}{a^3 \left(P_X + 2X P_{XX}\right)}\nonumber \\
&& G_{\varphi \varphi} = P_{\varphi\varphi} - \frac{2X\,P_{X\varphi}^2}{ \left(P_X + 2X P_{XX}\right)}
\end{eqnarray}

Using these definitions it is possible to invert all phase-space
quantities to the ones in configuration-space. To start with, we first
consider speed of sound. In conformal coordinate, it is given by
equation (\ref{sonic-conf}) and hence using above relations, we get
\begin{equation}
c_s^2 = \frac{P_X}{P_X + 2XP_{XX}}
\end{equation}
which matches with the conventional configuration-space definition of
sound of speed.  Similarly, the zeroth order equations
(\ref{ZeroHamP}), (\ref{ZeroEomaP}) and (\ref{ZeroEoMphiP}), after
inversion, become
\begin{eqnarray}
\label{ZeroHamP}
 && H^2 =- \frac{\kappa}{3}  (P_X {\varphi^{\prime}_0}^{2} + P a^2), ~~~ \mbox{Hubble  Constant:}~ H \equiv \frac{a^\prime}{a}    \\
 \label{ZeroEomaP}
&&- 2 \frac{a^{\prime \prime}}{a} +   H^2 = \kappa P a^2, \\
\label{ZeroEoMphiP}
&& P_X {\varphi_0^{\prime \prime}} - P_{XX} {\varphi_0^{\prime \prime}}
{\varphi_0^{\prime}}^{2} a^{-2} + P_{X\varphi}
{\varphi_0^{\prime}}^{2} + 2 P_X {\varphi_0^{\prime}} H +
P_{XX} H {\varphi_0^{\prime}}^{3} a^{-2} + P_{\varphi}
a^{2} = 0,
\end{eqnarray}
respectively. At first order, $N_1 = a \phi_1$ and $N^i = \delta^{i
  j} \partial_j B_1$, which helps to reduce first order perturbed
Hamiltonian constraint (\ref{FirstHamiltonian}) into
\begin{eqnarray}
&& \frac{H}{\kappa} \nabla^2 B_1 + \frac{3\, H^2}{\kappa} \phi_1 + \frac{G^2_{\pi_\varphi}}{2\,G_{\pi_\varphi \pi_\varphi}} \phi_1 a^2 + \frac{G_{\pi_\varphi}}{2\, G_{\pi_\varphi \pi_\varphi} a^2} \varphi_1^\prime + \frac{G_{\pi_\varphi}\,G_{\varphi \pi_\varphi}}{2\, G_{\pi_\varphi \pi_\varphi}} \varphi_1 a^2 \nonumber \\
&&~~~~~~~~ - \frac{G_\varphi}{2} \varphi_1 a^2 = 0
\end{eqnarray}

which, further, can be inverted back to configuration-space, again by
using above relations as
\begin{eqnarray}
\label{firstHamP}
&&\mathcal{H} \nabla^2{B_1} = \frac{\kappa}{2} \Big[ P_X \phi_1
{\varphi_0^{\prime}}^{2} + 2 P a^2 \phi_1 + 
P_X \varphi_0^{\prime} \varphi_1^{\prime} + P_{XX}
\phi_1 {\varphi_0^{\prime}}^{4} a^{-2} 
- P_{XX} \varphi_1^{\prime} {\varphi_0^{\prime}}^{3} a^{-2}
+ \nonumber \\
&& ~~~~~~~~~~~~~P_{X \varphi} {\varphi_0^{\prime}}^{2} \varphi_1 
+ P_\varphi \varphi_1 a^2 \Big].
\end{eqnarray}

Similarly, first order Momentum constraint becomes
\begin{equation}
\label{fisrtMomP}
\partial_i\phi_1= - \frac{\kappa}{2 \,H}P_X \varphi_0^{\prime} \partial_i\varphi_1
\end{equation}

and equation of motion of scalar field $\varphi_1$, i.e., equation (\ref{FirstEoMvarphi}) takes the form

\begin{eqnarray}
\label{FirstEoMP}
&&- P_X {\varphi_1^{\prime \prime}} a^{2} - P_{XX}
{\phi_1^{\prime}}{\varphi_0^{\prime}}^{3} + P_{XX} {\varphi_1^{\prime
    \prime}} {\varphi_0^{\prime}}^{2} - P_{XX\varphi} \phi_1
{\varphi_0^{\prime}}^{4} 
+ P_{XX\varphi} {\varphi_1^{\prime}}
{\varphi_0^{\prime}}^{3} -P_{\varphi} \phi_1 a^{4}- \nonumber\\
&& P_{\varphi \varphi} a^{4} \varphi_1 + P_X \phi_1 {\varphi_0^{\prime \prime}}
a^{2} + P_X \nabla^2{\varphi_1} a^{2} + P_X {\phi_1^{\prime}}{\varphi_0^{\prime}} a^{2} - 2 P_X {\varphi_1^{\prime}} H
a^2 - 4 P_{XX} \phi_1 {\varphi_0^{\prime
    \prime}}{\varphi_0^{\prime}}^{2} +\nonumber\\
  && 3 P_{XX} {\varphi_0^{\prime}}
{\varphi_1^{\prime}}{\varphi_0^{\prime \prime}} + P_{XX\varphi}
{\varphi_0^{\prime \prime}}{\varphi_0^{\prime}}^{2} \varphi_1 -
P_{X\varphi} {\varphi_0^{\prime}} {\varphi_1^{\prime}} a^{2} -
P_{X\varphi} {\varphi_0^{\prime \prime}} a^{2} \varphi_1 -P_{X\varphi\varphi} {\varphi_0^{\prime}}^{2} a^{2} \varphi_1 + \nonumber\\
&&  2 P_X
\phi_1 {\varphi_0^{\prime}} H a^2 + P_X {\varphi_0^{\prime}}
\nabla^2{B_1} a^{2} + P_{XX} \phi_1 H
{\varphi_0^{\prime}}^{3} - P_{XX} {\varphi_1^{\prime}} H
{\varphi_0^{\prime}}^{2} - P_{XXX} \phi_1 H
{\varphi_0^{\prime}}^{5} a^{-2} +\nonumber \\
&& P_{XXX} \phi_1 {\varphi_0^{\prime
    \prime}}{\varphi_0^{\prime}}^{4} a^{-2} + P_{XXX}
{\varphi_1^{\prime}}H{\varphi_0^{\prime}}^{4} a^{-2} -
P_{XXX}{\varphi_1^{\prime}}{\varphi_0^{\prime
    \prime}}{\varphi_0^{\prime}}^{3} a^{-2} - P_{XX\varphi}
H {\varphi_0^{\prime}}^{3} \varphi_1 - \nonumber\\
&&  2 P_{X\varphi}
{\varphi_0^{\prime}} H \varphi_1 a^2 = 0.
\end{eqnarray} 

Equations (\ref{ZeroHamP}), (\ref{ZeroEomaP}), (\ref{ZeroEoMphiP}),
(\ref{firstHamP}), (\ref{fisrtMomP}) and (\ref{FirstEoMP}) are
consistent with the zeroth and first order perturbed Euler-Lagrange
equations of motion.

\section{Interaction Hamiltonian}\label{InteractionHam}

The higher-order physical observables like Bi-spectrum/Tri-spectrum
are related to higher-order correlation functions; in order to compute
higher-order correlation functions, we need higher-order interaction
Hamiltonian. In this section, we obtain the interaction Hamiltonian of
the non-canonical field. Third order perturbed generalized interaction
Hamiltonian for non-canonical scalar field in terms of phase-space
variables is obtained directly by expanding the Hamiltonian
(\ref{Hamfull}) upto third order of perturbation \cite{Nandi:2015ogk}
and it takes the form
\begin{eqnarray}
\label{ThirdIntHam}
\mathcal{H}_3 &=&  - N_1 {\delta}_{i j} {\delta}_{k l} \kappa {\pi_1}^{i j} {\pi_1}^{k l} a + 2\, N_1 {\delta}_{i j} {\delta}_{k l} \kappa {\pi_1}^{i k} {\pi_1}^{j l} a - \frac{1}{2}\, G_{\varphi\varphi} N_1 \varphi_1{}^{2} a{}^{3} - \frac{1}{2}\, G_{\pi_\varphi \pi_\varphi} N_1 \pi_{\varphi1}{}^{2} a{}^{3} -\nonumber \\
&& G_{\varphi\pi_\varphi} N_1 \pi_{\varphi1} a{}^{3} \varphi_1 - G_Y N_1 {\delta}^{i j} {\partial}_{i}{\varphi_1}\,  {\partial}_{j}{\varphi_1}\,  a - G_{Y \pi_\varphi} N_0 \pi_{\varphi1} {\delta}^{i j} {\partial}_{i}{\varphi_1}\,  {\partial}_{j}{\varphi_1}\,  a -\nonumber \\
&& G_{\varphi Y} N_0 {\delta}^{i j} {\partial}_{i}{\varphi_1}\,  {\partial}_{j}{\varphi_1}\,  \varphi_1 a - \frac{1}{6}\, G_{\pi_\varphi \pi_\varphi \pi_\varphi} N_0 \pi_{\varphi1}{}^{3} a{}^{3} - \frac{1}{6}\, G_{\varphi \varphi \varphi} N_0 \varphi_1{}^{3} a{}^{3} - \nonumber \\
&&\frac{1}{2}\, G_{\varphi \pi_\varphi \pi_\varphi} N_0 \pi_{\varphi1}{}^{2} a{}^{3} \varphi_1 - \frac{1}{2}\, G_{\varphi\varphi\pi_\varphi} N_0 \pi_{\varphi1} \varphi_1{}^{2} a{}^{3} + {N_1}^{i} \pi_{\varphi1} {\partial}_{i}{\varphi_1}
\end{eqnarray}
and similarly, fourth order interaction Hamiltonian takes the form
\begin{eqnarray}
\label{FourthIntHam}
\mathcal{H}_4 &=&  - \frac{1}{2}\, G_{YY} N_0 {\delta}^{i j} {\delta}^{k l} {\partial}_{i}{\varphi_1}\,  {\partial}_{j}{\varphi_1}\,  {\partial}_{k}{\varphi_1}\,  {\partial}_{l}{\varphi_1}\,  a{}^{-1} - G_{\pi_\varphi Y} N_1 \pi_{\varphi1} {\delta}^{i j} {\partial}_{i}{\varphi_1}\,  {\partial}_{j}{\varphi_1}\,  a - \nonumber \\
&& G_{\varphi Y} N_1 {\delta}^{i j} {\partial}_{i}{\varphi_1}\,  {\partial}_{j}{\varphi_1}\,  \varphi_1 a - \frac{1}{6}\, G_{\pi_\varphi \pi_\varphi \pi_\varphi} N_1 \pi_{\varphi1}{}^{3} a{}^{3} - \frac{1}{6}\, G_{\varphi\varphi\varphi} N_1 \varphi_1{}^{3} a{}^{3} - \nonumber \\
&&\frac{1}{2}\, G_{\varphi \pi_\varphi \pi_\varphi} N_1 \pi_{\varphi1}{}^{2} a{}^{3} \varphi_1 - \frac{1}{2}\, G_{\pi_\varphi \pi_\varphi Y} N_0 {\delta}^{i j} {\partial}_{i}{\varphi_1}\,  {\partial}_{j}{\varphi_1}\,  \pi_{\varphi1}{}^{2} a - \frac{1}{2}\, G_{\varphi\varphi \pi_\varphi} N_1 \pi_{\varphi1} \varphi_1{}^{2} a{}^{3} - \nonumber \\
&&\frac{1}{2}\, G_{\varphi\varphi Y} N_0 {\delta}^{i j} {\partial}_{i}{\varphi_1}\,  {\partial}_{j}{\varphi_1}\,  \varphi1{}^{2} a - G_{\varphi \pi_\varphi Y} N_0 \pi_{\varphi1} {\delta}^{i j} {\partial}_{i}{\varphi_1}\,  {\partial}_{j}{\varphi_1}\,  \varphi_1 a - \nonumber \\
&&\frac{1}{24}\, G_{\pi_\varphi \pi_\varphi \pi_\varphi \pi_\varphi} N_0 \pi_{\varphi1}{}^{4} a{}^{3} - \frac{1}{24}\, G_{\varphi\varphi\varphi\varphi} N_0 \varphi_1{}^{4} a{}^{3} - \frac{1}{6}\, G_{\varphi\pi_\varphi\pi_\varphi\pi_\varphi} N_0 \pi_{\varphi1}{}^{3} a{}^{3} \varphi_1 - \nonumber \\
&&\frac{1}{6}\, G_{\varphi\varphi\varphi\pi_\varphi} N_0 \pi_{\varphi1} \varphi_1{}^{3} a{}^{3} - \frac{1}{4}\, G_{\varphi\varphi\pi_\varphi\pi_\varphi} N_0 \pi_{\varphi1}{}^{2} \varphi_1{}^{2} a{}^{3}.
\end{eqnarray}

Again, using inversion formulae mentioned in above section,
phase-space form of interaction Hamiltonian can be written in terms of
configuration-space variables.

\section{Extension to generalized higher-derivative models}\label{ExGal}

As we have shown above, for an arbitrary non-canonical scalar field,
it is possible to define a canonical conjugate momenta and the
Hamiltonian. In this section, we extend our method to generalize higher-derivative models. First, we extend the analysis to 
  G-Inflation\cite{Kobayashi2010, Kobayashi2011} model with
  generalized functions $P(X, \varphi)$ and $K(X, \varphi)$ .
  
Action for G-Inflation scalar field minimally coupled to gravity is
given by

\begin{equation}\label{galaction}
  \mathcal{S}_G = \int d^4 x \sqrt{-g} \left[\frac{1}{2 \kappa} R + P(X, \varphi) + K(X, \varphi) \Box \varphi \right].
\end{equation}

Directly obtaining the Hamiltonian for the above action is difficult
since it contains second order derivatives of the scalar
field. However, using the approach of Deffayet \emph{et al.}
\cite{Deffayet:2015qwa, Nandi:2015ogk}, action (\ref{galaction}) can
be re-written as
\begin{eqnarray}\label{GalileanAction}
\mathcal{S}_G &=& \int d^4 x \sqrt{-g} \left[\frac{1}{2 \kappa} R + P(X, \varphi) + K(X, \varphi)\, S \right] + \int d^4 x \,\lambda\, (S - \Box \varphi) \nonumber \\
&=& \int d^4 x \sqrt{-g} \left[\frac{1}{2 \kappa} R + P(X, \varphi) + K(X, \varphi)\, S \right]+ \int d^4x [\, \lambda S + \lambda~ g^{\mu \nu}~\Gamma^{\alpha}_{\mu \nu} ~\partial_{\alpha}{\varphi} + \nonumber \\
&& ~~~~~~~~g^{\mu \nu} \partial_{\mu}{\varphi} \partial_{\nu}\lambda + \lambda \, \partial_{\nu}{g^{\mu \nu}} \partial_{\mu}{\varphi}\,].
\end{eqnarray}

Linearizing the action costs two extra variables in
configuration-space, thus four extra phase-space variables. We have
discussed the issue in I \cite{Nandi:2015ogk} and proved that those
variables are not dynamic in nature and thus, there are no extra
degrees of freedom.

Since the action (\ref{GalileanAction}) is converted in terms of first
derivatives of fields, it is now possible to define momenta in terms
of time derivative of the fields. However, the action still contains
two generalized configuration-space variables $P(X, \varphi)$ and
$G(X, \varphi)$. Hence, using the above approach for generalized
non-canonical scalar field, a consistent perturbed Hamiltonian
formalism for generalized Galilean scalar field can be established.

 The approach can also be extended to any other
  higher-derivative models like Hordenski scalar field models,
  modified gravity models or an arbitrary higher-derivative
  theory. The above case is the quadratic and cubic parts of the
  Hordenski's scalar field model\cite{Horndeski1974}. In case of general Hordenski's
  scalar field model, action depends on $R_{\mu \nu}, \nabla_{\mu
    \nu}\varphi, \partial_{\mu}{\varphi}, g_{\mu \nu}$ and
  $\varphi$. The metric-part can be written in terms of extrinsic
  tensor, $K_{i j}$ and 3-Ricci scalar, $^{(3)}R$ with Lapse function,
  $N$ and Shift vector, $N^i$. Since the action contains
  $\nabla_{\mu \nu} \varphi$ instead of only $\Box \varphi$, above
  method for linearizing action will not work. Instead, we have to
  linearize the action by adding $$ S_H + \int d^4x~\lambda^{\mu
    \nu}\left(S_{\mu \nu} - \nabla_{\mu \nu}\varphi\right) $$ for
  general Hordenski's model \cite{Deffayet:2015qwa}. Hordenski's full
  action contains four unknown functions, $G_n(X, \varphi),\,
  n=2\cdots5$, hence, using the approach for non-canonical
  scalar field, we can also deal with Hamiltonian formulation for
  Hordenski's theory.

  By using the same argument and method, it is possible to obtain
  Hamiltonian beyond Hordenski's model, i.e., for any higher-order
  derivative gravity models with arbitrary functions. To deal with
  arbitrary functions, using the above approach for generalized
  non-canonical scalar fields, we can define a new and unique
  phase-space variable(s) and write the corresponding canonical
  Hamiltonian of the system and inversion formulae can be used to
  invert from phase-space variable to configuration-space variable and
  vice-versa. Once the Hamiltonian is obtained, we can use the
  approach in I to get the consistent Hamiltonian formalism of
  cosmological perturbation at any perturbed order for the specific
  model.

  Hamiltonian approach in I, is independent of how we construct the
  Hamiltonian and is readily applicable once we successfully write
  down a consistent Hamiltonian for a specific model. Hence, the
  Hamiltonian approach for higher derivative theory is not restricted
  only by the Deffayet's approach \cite{Deffayet:2015qwa}. Recently,
  Langlois and Noui \cite{Langlois:2015cwa, Langlois:2015skt} have
  also provided a simpler way to obtain Hamiltonian for higher
  derivative theory and the Hamiltonian approach for perturbation can
  also be extended to these models. 

\section{Conclusion and discussion}\label{Conclu}

In this work, we have explicitly provided the Hamiltonian formulation
of cosmological perturbation theory for generalized non-canonical
scalar fields. The following procedure was adopted: first we provided
the essential information regarding gauge-choices and related
gauge-invariant quantities. Next, we performed Legendre transformation
for the generalized non-canonical scalar fields and showed that, since
($\varphi^\prime \rightarrow \pi_\varphi$) transformation is not
possible, Hamiltonian for generalized non-canonical scalar fields
cannot be obtained by using conventional method.

We introduced a new generalized phase-space variable $G(\pi_\varphi,
\gamma, Y, \varphi)$ that is unique for different non-canonical scalar
fields and obtained Hamiltonian of a non-canonical scalar field. We
showed that, this is the simplest and efficient way to obtain the
Hamiltonian. We extended the approach in I to generalized
non-canonical scalar fields in the flat-slicing that doesn't lead to
gauge-artifacts and obtained perturbed Hamilton's equations in terms
of phase-space variables. In parallel, we also extended Langlois'
approach to generalized non-canonical scalar field and showed that
both approaches lead to identical speed of sound.

In order to compare Hamiltonian approach with Lagrangian approach,
Hamilton's equations are to be converted to Euler-Lagrange equation
and in doing so, we provided explicit forms of $G(\pi_\varphi, \gamma,
Y, \varphi)$ for different non-canonical scalar field models and
showed that the Hamiltonian formulation is consistent.

Since we do not know how, in general, phase-space derivatives of
$G(\pi_\varphi, \gamma, Y, \varphi)$ transform to configuration-space
derivatives, hence for an arbitrary field, it is not possible to
directly invert the generalized phase-space Hamilton's equations to
Euler-Lagrange equations. In order to overcome this, we prescribed an
inversion mechanism from generalized phase-space variables to
generalized configuration-space variables (and vice versa) and showed
that all generalized phase-space equations lead to consistent E-L
equations. We also retrieved the conventional form of speed of sound
in configuration-space.

We also obtained the interaction Hamiltonian in terms of phase-space
variables for generalized non-canonical scalar field at third and
fourth order of perturbation for scalar perturbations. These can also
be expressed in terms of $(\varphi^\prime,~ \varphi)$ using the
general inversion formulae. Note that, we considered only the first
order scalar perturbations. Vector or tensor modes can similarly be
implemented by considering $\delta\gamma_{i j} \neq 0$ and decomposing
the metric using vector and tensor modes. Hamiltonian as well as
equations of motions for vector or tensor modes also change
accordingly. For the linear order, three modes decouple and
$\delta\pi^{i j}$ can also be decomposed as ${\delta\pi_{S}^{i j} +
  \delta\pi_{V}^{ij} + \delta\pi_{T}^{i j}}$, so the equations of
motion. However, for higher order of perturbations, modes are highly
coupled to each other, hence similar decomposition is not possible.

Finally, we briefly discussed the Hamiltonian formulation for
generalized higher derivative scalar fields. The method
is not restricted to gravity related models, it can also be applied to
any other models where the Lagrangian is not specified properly.

Throughout the work, we carried out the method by assuming that the
field allows Legendre transformation, which, most of the known models
follow. However, if a certain model is specified in such a way that
$\varphi^\prime$ cannot be written in terms of $\pi_\varphi$ or the
mapping is one-to-many, then the current formalism cannot be applied
to obtain a unique form of $G(\pi_\varphi, \gamma, Y, \varphi)$ and
hence, for those kind of models, this approach is not applicable.

\section{Acknowledgements}
We thank Swastik Bhattacharya for useful discussions. Work is
supported by Max Planck-India Partner group on Gravity and Cosmology.
DN is supported by CSIR fellowship. Further we thank Kasper Peeters
for his useful program
Cadabra\cite{Peeters:2007wn,DBLP:journals/corr/abs-cs-0608005} and
useful algebraic calculations with it.

\appendix

\section{Inversion formulae of $X$ and $\pi_\varphi$ and $G(\gamma,
  \pi_\varphi, Y, \varphi)$ for different scalar field
  models}{\label{Inversion}}
\subsection{Canonical scalar field}{\label{InvCanonical}}
In case of canonical scalar field, $P(X, \varphi)$ is given by $$P(X, \varphi) = - X - V(\varphi).$$
Hence, using equation (\ref{piphigeneral}), we get
\begin{eqnarray}
\pi_\varphi &=& \sqrt{\gamma}\sqrt{- 2X + Y} \nonumber \\
\Rightarrow X &=& \frac{1}{2} Y - \frac{\pi_\varphi^2}{2 \gamma}
\end{eqnarray}
and $G(\gamma, \pi_\varphi, Y, \varphi)$ is given by the relation (\ref{G})
\begin{equation}
G(\gamma, \pi_\varphi, Y, \varphi) = -\frac{1}{2}\frac{\pi_\varphi^2}{\gamma} - \frac{1}{2}\,Y - V(\varphi).
\end{equation}

\subsection{Tachyonic field}
Tachyons are described by $$P(X, \varphi) = - V(\varphi) \sqrt{1 + 2X}.$$
Similarly, in case of Tachyons, we get
\begin{eqnarray}
X &=& \frac{\gamma \, V^2\, Y - \pi_\varphi^2}{2\,(\pi_\varphi^2 + \gamma\, V^2)} \\
G(\gamma, \pi_\varphi, Y, \varphi) &=& -\frac{1}{\sqrt{\gamma}}\,\sqrt{1 + Y} \sqrt{\pi_\varphi^2 + \gamma \,V^2}.
\end{eqnarray}

\subsection{DBI field}
For DBI field, $$P(X, \varphi) = -\frac{1}{f(\varphi)} \left(\sqrt{1 +
    2\,f(\varphi)\, X} - 1 \right)- V(\varphi)$$ which implies that,
\begin{eqnarray}
X &=& \frac{\gamma \, Y - \pi_\varphi^2}{2\,(\gamma + f\,\pi_\varphi^2 )} \\
G(\gamma, \pi_\varphi, Y, \varphi) &=& -\frac{1}{\gamma\,f(\varphi)}\sqrt{\left(\gamma + f(\varphi)\, Y\right) \left(f(\varphi)\, \pi_\varphi^2 + \gamma\right)} + \frac{1}{f(\varphi)} - V(\varphi)
\end{eqnarray}

\section{Langlois' approach for non-canonical scalar field}\label{Langlois-nonC}
Two decades back, Langlois' obtained a consistent Hamiltonian for
canonical scalar field\cite{Langlois1994}. In this section, we extend
the method to non-canonical scalar fields.

Following \cite{Langlois1994}, expressing background 3-metric
$\gamma_{0 i j} = e^{2 \alpha}$, it can be shown that the first order
perturbed Hamiltonian constraint takes the form

\begin{eqnarray}\label{LangHam}
  \mathcal{H}_{N1} &\equiv & - \frac{e^{3 \alpha}}{2 \kappa}\, \left[\,\gamma_{0}^{i k} \gamma_0^{j l} - \gamma_{0}^{i j} \gamma_0^{k l} \,\right] \partial_{i j} \gamma_{1 k l} - \frac{\kappa}{3} e^{- 3 \alpha} \,\pi_\alpha\, \gamma_{0ij}\, \pi_1^{i j} - \Big[\,\frac{\kappa}{72}\,e^{-3\alpha}\,\pi_\alpha^2 + \nonumber \\
  &&\frac{1}{2}\,e^{3\alpha}\, G \,\Big]\,\gamma_0^{i j}\, \gamma_{1 i j} - e^{3\alpha}\, G_{\pi_\varphi}\, \pi_{\varphi 1} - e^{3\alpha}\,G_{\varphi}\, \varphi_1 = 0,
\end{eqnarray}  
where $\pi_\alpha$ is the momentum corresponding to $\alpha$. Similarly, first order perturbed Momentum constraint becomes
\begin{equation}\label{LangMom}
\mathcal{H}_{1i} \equiv - 2\, \partial_k \gamma_{1i j}\, \pi_0^{j k} - 2\, \gamma_{0ij}\, \partial_k \pi_1^{j k} + \pi_0^{j k}\, \partial_i \gamma_{1jk} + \pi_{\varphi1}\, \partial_i \varphi_1 = 0.
\end{equation}
 
In momentum space, equations (\ref{LangHam}) and (\ref{LangMom}) becomes
\begin{eqnarray}\label{HamK}
  \mathcal{H}_{N1}(k) &\equiv & - \frac{e^{3 \alpha}}{2 \kappa}\, \left[\,\gamma_{0}^{i k} \gamma_0^{j l} - \gamma_{0}^{i j} \gamma_0^{k l} \,\right] k_i k_j \gamma_{1 k l} - \frac{\kappa}{3} e^{- 3 \alpha} \,\pi_\alpha\, \gamma_{0ij}\, \pi_1^{i j} - \Big[\,\frac{\kappa}{72}\,e^{-3\alpha}\,\pi_\alpha^2 + \nonumber \\
  &&\frac{1}{2}\,e^{3\alpha}\, G \,\Big]\,\gamma_0^{i j}\, \gamma_{1 i j} - e^{3\alpha}\, G_{\pi_\varphi}\, \pi_{\varphi 1} - e^{3\alpha}\,G_{\varphi}\, \varphi_1 = 0 \\
\label{MomK}
\mathcal{H}_{1i}(k) &\equiv & - 2\, k_k\, \gamma_{1i j}\, \pi_0^{j k} - 2\, \gamma_{0ij}\, k_k\, \pi_1^{j k} + \pi_0^{j k}\, k_i\, \gamma_{1jk} + \pi_{\varphi1}\, k_i\, \varphi_1 = 0
\end{eqnarray}

The scalar configuration variables are
\begin{equation}
  \gamma_1  = \frac{1}{3} \gamma_0^{i j} \, \gamma_{1ij}, \quad \gamma_2 = \frac{1}{2}\,\left[\frac{3 k^i \,k^j}{k^2}\,\gamma_{1ij} - \gamma_0^{i j} \, \gamma_{1ij}\right],
\end{equation}
which are associated with their conjugate momenta
\begin{equation}
  \pi^1 = \gamma_{0ij}\, \pi_1^{i j}, \quad \pi^2 = \frac{k_i k_j}{k^2}\, \pi_1^{i j} - \frac{1}{3} \gamma_{0ij} \pi_1^{i j}.
\end{equation}

Hence the energy constraint (\ref{HamK}) becomes
\begin{eqnarray}
  E &=& - \left[ \frac{1}{24} \kappa e^{-3\alpha}\, \pi_\alpha^2 + \frac{3}{2} e^{3\alpha}\, G\right]\, \gamma_1 - \frac{e^{3\alpha}}{\kappa}\,k^2 \,\gamma_1 + \frac{e^{3\alpha}}{3 \kappa}\, k^2 \, \gamma_2 \nonumber \\
  && -\frac{\kappa}{3} e^{-3\alpha}\, \pi_\alpha\, \pi^1 - e^{3\alpha}\, G_{\pi_\varphi}\, \pi_{\varphi1} - e^{3\alpha}\, G_{\varphi}\, \varphi_1 = 0.
\end{eqnarray}

Momentum constraint contains scalar and vector, both
modes. Contracting with $k^i$, we obtain the scalar part of Momentum
constraint
\begin{equation}
M \equiv \frac{1}{6} \pi_\alpha\, \gamma_1 - \frac{2}{9} \pi_\alpha \, \gamma_2 - \frac{2}{3}\pi^1 - 2 \pi^2 + \pi_\alpha\,\varphi_1 = 0
\end{equation}

In case of scalar phase-space, there exist two first class constraints, namely $E$ and $M$ and
\begin{eqnarray}\label{ham1}
E\left(\gamma_\alpha, \pi^\beta = \frac{\partial S}{\partial \gamma_\beta}\right) &=& 0 \\
\label{mom1}
M\left(\gamma_\alpha, \pi^\beta = \frac{\partial S}{\partial \gamma_\beta}\right) &=& 0
\end{eqnarray}
where $S$ is the quadratic generating function, given by
\begin{equation}
S = 	\frac{1}{2}A_{\alpha\beta}\, \gamma_\alpha\,\gamma_\beta+ B_\alpha\,\gamma_\alpha.
\end{equation}
$\alpha, \beta = 0, 1, 2, \gamma_0 = \varphi_1, \pi^0 =
\pi_{\varphi1}$ and $A_{\alpha\beta}$ are symmetric. Hence, equations
(\ref{ham1}) and (\ref{mom1}) become equations for $A_{\alpha\beta}$
and $B_\alpha$ with a polynomial form in $\gamma_\alpha$ and lead to
the following four equations for the Energy constraint:
\begin{eqnarray}
  &&- \left[\frac{1}{24}\kappa\, e^{-3\alpha}\, \pi_\alpha^2 + \frac{3}{2}e^{3\alpha} \,G\right] - \frac{e^{3\alpha}}{\kappa} k^2 - \frac{e^{-3\alpha}\, \kappa}{3}\pi_\alpha\, A_{11} - e^{3\alpha}\, G_{\pi_\varphi} \,A_{01} = 0 \\
  && \frac{e^{3\alpha}}{3\kappa}k^2 - \frac{\kappa}{3} e^{-3\alpha}\, \pi_\varphi \, A_{12} - e^{3\alpha}\,G_{\pi_\varphi}\,A_{02} = 0 \\
  && -\frac{\kappa}{3} e^{-3\alpha}\,\pi_\alpha\, A_{01} - e^{-3\alpha}\, G_{\pi_\varphi}\,A_{00}- e^{3\alpha}\,G_{\varphi} = 0 \\
  && -\frac{\kappa}{3} e^{-3\alpha}\, \pi_\alpha\,B_1- e^{3\alpha}\,G_{\pi_\varphi}\, B_0 = 0
\end{eqnarray}
 and for the Momentum constraint
\begin{eqnarray}
&& \frac{1}{6} \pi_\alpha - \frac{2}{3} A_{11} - 2 A_{21} = 0\\
&& -\frac{2}{9}\pi_\alpha - \frac{2}{3} A_{12}- 2 A_{22} = 0 \\
&& -\frac{2}{3}A_{10} - 2 A_{20} + \pi_\alpha = 0 \\
&& -\frac{2}{3}B_1 - 2 B_2 = 0
\end{eqnarray}
respectively.

The solutions for the $B_\alpha$ form a one-dimensional space and can be written as
\begin{equation}
B_0 = P, \quad B_1 = -\frac{3}{\kappa} e^{6\alpha}\,\frac{G_{\pi_\varphi}}{\pi_\alpha}\,B_0, \quad B_2 = \frac{e^{6\alpha}}{\kappa}\, \frac{G_{\pi_\varphi}}{\pi_\alpha}\,B_0
\end{equation}
where dependence of the $B_\alpha$ on the free parameter $P$ is chosen
for later convenience. $A_{\alpha\beta}$ are undetermined since there
are five out of six independent equations and one is background
equation. The additional condition is arbitrary and independent of any
physical change in the system. The quantity $P$ is the momentum in the
reduced phase-space and its conjugate coordinate is given by

\begin{equation}
Q = \frac{\partial S}{\partial P} =  \varphi_1 + \frac{e^{6\alpha}}{\kappa}\,\frac{G_{\pi_\varphi}}{\pi_\alpha}\, \left(\gamma_2 - 3\gamma_1\right)
\end{equation}
which coincides with gauge-invariant Mukhanov's variable. Other relations between old and new variables are given as
\begin{eqnarray}
\varphi_1 &=& Q + [\gamma_1, \gamma_2], \quad \pi_{\varphi1} = A_{00}\,Q + P + [\gamma_1, \gamma_2] \\
\pi^1 &=& A_{10}\,Q - \frac{3}{\kappa}\,e^{6\alpha}\,\frac{G_{\pi_\varphi}}{\pi_\alpha}\,P + [\gamma_1, \gamma_2], \quad \pi^2 = A_{20}\, Q + \frac{e^{6\alpha}}{\kappa}\,\frac{G_{\pi_\varphi}}{\pi_\alpha}\, P + [\gamma_1, \gamma_2]
\end{eqnarray}
where brackets contain all the terms with $\gamma_1$ or
$\gamma_2$. These are not written explicitly since they are `pure
gauge' and do not contribute to the `true' dynamics.

The second order expansion of the Energy constraint is given by
\begin{eqnarray}\label{b24}
  \mathcal{H}_{N2} &=& \frac{2\kappa}{\sqrt{\gamma}}\, \left(\gamma_{0ik} \gamma_{0jl} -\frac{1}{2} \gamma_{0ij} \gamma_{0kl}\right)\,\pi_1^{ij}\pi_1^{k l} - \frac{\sqrt{\gamma}}{2}\, G_{\pi_\varphi\pi_\varphi}\,\pi_{\varphi1}^2 \nonumber \\
  &&-\frac{\sqrt{\gamma}}{2}\,G_{\varphi\varphi}\,\varphi_1^2-\sqrt{\gamma}\,G_Y\,\gamma_0^{i j} \partial_i\varphi_1\partial_j \varphi_1-\sqrt{\gamma}\,G_{\varphi\pi_\varphi}\,\pi_{\varphi1}\,\varphi_1 + [\gamma_{i j}]
\end{eqnarray}
where $[\gamma_{i j}]$ collectively represents all the terms that
involve $\gamma_{1 i j}$. By choosing $N=1$ and $N^i = 0$ to simplify
calculations, the scalar part of the Hamiltonian is easily obtained
and is given by
\begin{eqnarray}\label{b25}
  H^s &=& \int d^3k \left\{N \mathcal{H}_{N} + N^i \mathcal{H}_i\right\} \nonumber \\
  &=& \int d^3k \Big\{\frac{2\kappa}{\sqrt{\gamma}}\,\left(-\frac{1}{6}\pi_1{}^2 - \frac{3}{2}\pi_2{}^2\right) - \frac{\sqrt{\gamma}}{2}\, G_{\pi_\varphi\pi_\varphi}\,\pi_{\varphi1}^2 \nonumber \\
  &&-\frac{\sqrt{\gamma}}{2}\,G_{\varphi\varphi}\,\varphi_1^2-\sqrt{\gamma}\,G_Y\,k^2 \varphi_1^2-\sqrt{\gamma}\,G_{\varphi\pi_\varphi}\,\pi_{\varphi1}\,\varphi_1 + [\gamma_1, \gamma_2]
  \Big\}.
\end{eqnarray}
Hence the gauge-invariant Hamiltonian is given by
\begin{eqnarray}
  H_{GI}^s &=& H^s + \{S, H_0\}_{Background} \\
  &=& \int d^3k\Bigg[ -\frac{\sqrt{\gamma}}{2}\,G_{\pi_\varphi\pi_\varphi}\,P^2 + \Big\{-\frac{\kappa}{3\sqrt{\gamma}}\,A_{10}^2 + \frac{3\kappa}{\sqrt{\gamma}}\,A_{20}^2 - \frac{\sqrt{\gamma}}{2}\,G_{\pi_\varphi\pi_\varphi}\,A_{00}^2 \nonumber \\
  && -\frac{\sqrt{\gamma}}{2}\,G_{\varphi\varphi}+ \sqrt{\gamma}\,G_Y\,k^2 + \frac{1}{2}\,\dot{A}_{00} - \sqrt{\gamma}\,G_{\varphi\pi_\varphi}\,A_{00}\Big\}Q^2 \nonumber \\
  && + \left\{\frac{2}{\sqrt{\gamma}}e^{6\alpha}\,\frac{G_{\pi_\varphi}}{\pi_\alpha}\,A_{10} + \frac{6}{\sqrt{\gamma}}\, e^{6\alpha}\, \frac{G_{\pi_\varphi}}{\pi_\alpha}\, A_{20} - \sqrt{\gamma}\, G_{\pi_\varphi\pi_\varphi}\,A_{00}\right\}\,P\,Q\Bigg]
\end{eqnarray}
where $\dot{A}_{00} = \{A_{00}, H_0\}_{Background}$. If we impose additional condition
\begin{equation}
\frac{2}{\sqrt{\gamma}}e^{6\alpha}\,\frac{G_{\pi_\varphi}}{\pi_\alpha}\,A_{10} + \frac{6}{\sqrt{\gamma}}\, e^{6\alpha}\, \frac{G_{\pi_\varphi}}{\pi_\alpha}\, A_{20} - \sqrt{\gamma}\, G_{\pi_\varphi\pi_\varphi}\,A_{00} = 0
\end{equation}
in order to cancel cross terms in the above Hamiltonian, we get the following solutions:
\begin{eqnarray}
A_{00} &=& \frac{3\, G_{\pi_\varphi}}{G_{\pi_\varphi\pi_\varphi}}\,\frac{\pi_{\varphi0}}{\pi_\alpha}, \quad A_{10} = -\frac{9 }{\kappa}\,e^{6\alpha}\, \frac{G_{\pi_\varphi}^2\,\pi_{\varphi0}}{G_{\pi_\varphi \pi_\varphi}\,\pi_\alpha^2} - \frac{3}{\kappa}\,e^{6\alpha}\,\frac{\pi_{\varphi0}}{\pi_\alpha}, \nonumber \\
 A_{20} &=& \frac{3}{\kappa}\,e^{6\alpha}\,\frac{G_{\pi_\varphi}^2\,\pi_{\varphi0}}{G_{\pi_\varphi \pi_\varphi}\,\pi_\alpha^2} +  \frac{1}{\kappa} \,e^{6\alpha}\,\frac{\pi_{\varphi0}}{\pi_\alpha} + \frac{1}{2}\, \pi_{\varphi0}.
\end{eqnarray}
Finally, the Hamiltonian takes the form
\begin{eqnarray}
H_{GI}^{s} &=& \int d^3 k \left\{-\frac{1}{2}\,e^{3\alpha}\,G_{\pi_\varphi\pi_\varphi}\,P^2 + \frac{1}{2}\, \left(X + 2\, e^{3\alpha}\,k^2\,G_Y\right)\,Q^2 \right\}, \quad \mbox{where} \\
X &\equiv & 9\,e^{3\alpha}\,\frac{G_{\pi_\varphi}{}^2\,\pi_{\varphi0}{}^2}{G_{\pi_\varphi\pi_\varphi}\pi_\alpha{}^2} - 6\,e^{3\alpha}\,\frac{G_{\varphi\pi_\varphi}\,G_{\pi_\varphi}\,\pi_{\varphi0}}{\pi_\alpha} + \dot{A}_{00} + \frac{3}{2}\,\kappa\,e^{-3\alpha}\,\pi_{\varphi0}^2 \nonumber \\
&&\qquad \qquad 6 \,e^{3\alpha}\,\frac{G_{\varphi}\,\pi_{\varphi0}}{\pi_\alpha} - e^{3\alpha}\,G_{\varphi\varphi} 
\end{eqnarray}
and the corresponding equation of motion becomes
\begin{eqnarray}
\ddot{Q} - 2\,k^2\, e^{6\alpha}\, G_{\pi_\varphi\pi_\varphi}\, G_Y\,Q + \left(3\,H - \frac{\dot{G}_{\pi_\varphi\pi_\varphi}}{G_{\pi_\varphi\pi_\varphi}}\right)\,\dot{Q} - e^{3\alpha}\,G_{\pi_\varphi\pi_\varphi}\,X\,Q = 0.
\end{eqnarray}

Note that, speed of sound
\begin{eqnarray}
c_s^2 &=& 2\, e^{6\alpha}\, G_{\pi_\varphi\pi_\varphi}\, G_Y \nonumber \\
&=& 2\, a^6\, G_{\pi_\varphi\pi_\varphi}\, G_Y.
\end{eqnarray}

Note that, in Langlois' approach, the time coordinate represents
cosmic time and hence the sound speed, according to our approach, is
given by $2\, a^4\, G_{\pi_\varphi\pi_\varphi}\, G_Y$. The discrepancy
arises due to the fact that, in Langlois' approach, $\gamma^{i
  j} \partial_{i j} \rightarrow - k^2$ (see eqs. (\ref{b24}) and
(\ref{b25})), where we have used $\delta^{i j} \partial_{i j}
\rightarrow - k^2$. Hence, extra $a^2$ factor appears in Langlois'
approach.

%\bibliographystyle{JHEP}
%\bibliography{Mycollection}{}
\providecommand{\href}[2]{#2}\begingroup\raggedright\endgroup

\end{document}